# A New Way to Set the Energy Scale in Lattice Gauge Theories and its Application to the Static Force and $\alpha_s$ in SU(2) Yang–Mills Theory

R. Sommer
Deutsches Elektronen-Synchrotron DESY, Hamburg

**Abstract**

We introduce a hadronic scale $R_0$ through the force $F(r)$ between static quarks at intermediate distances $r$. The definition $F(R_0)R_0^2 = 1.65$ amounts to $R_0 \simeq 0.5$ fm in phenomenological potential models. Since $R_0$ is well defined and can be calculated accurately in a Monte Carlo simulation, it is an ideal quantity to set the scale. In SU(2) pure gauge theory, we use new data (and $R_0$ to set the scale) to extrapolate $F(r)$ to the continuum limit for distances $r = 0.18$ fm to $r = 1.1$ fm.

Through $R_0$ we determine the energy scale in the recently calculated running coupling, which used the recursive finite size technique to reach large energy scales. Also in this case, the lattice data can be extrapolated to the continuum limit. The use of one loop Symanzik improvement is seen to reduce the lattice spacing dependence significantly.

## 1. Introduction

In the limit when the mass of a quark $h$ becomes large compared to typical QCD–scales (such as the proton's mass), bound states $\bar{h}h$ are expected to be described by an effective nonrelativistic Schrödinger equation[1]. The nonrelativistic potential is given by the energy of static fundamental charges separated a distance $r$. In the real world, such a description should be approximately valid for the $\bar{b}b$ and maybe even for the $\bar{c}c$ spectra.

In fact, both these spectra can be described by *one effective* potential. Since (as one knows a posteriori from a successful model) the $\bar{b}b$ states have $rms$–radii of 0.2 fm to 0.7 fm and the $\bar{c}c$ states have a size of 0.4 fm to 1.0 fm, the spectra determine the *effective* potential in the range of $r \simeq 0.2$ fm to $r \simeq 1$ fm. Although we must remember that the relationship between the static QCD potential and the effective potential used in phenomenology is *not* well understood, this discussion suggests that the distance range where we have the best information on the force $F(r)$ between static quarks is at distances of around 0.5 fm.

In lattice gauge theory calculations, we need to fix one dimensionful quantity in order to set the overall scale. In the pure gauge theory, this is usually done through the string tension, defined as $K = \lim_{r \to \infty} F(r)$. The limiting procedure is not easy to do because the statistical and systematic errors on the force increase with the distance. It is therefore superior to use the force at intermediate distances to determine the scale: one calculates $R(c)$ such that $F(R(c))R(c)^2 = c$. Because of the above mentioned arguments, we chose $c = 1.65$, which corresponds to $R_0 \equiv R(1.65) = 0.49$ fm in the Cornell[2] and the Richardson[3] model and very similar values in the other successfull models.

The statistical and the systematic errors of $R_0$ are quite moderate. This opens the possibility of extrapolations to the continuum limit.

In section 3, we determine the force in the continuum limit of the SU(2) Yang Mills theory: the dimensionless quantity $H(x) = F(r)r^2|_{r=xR_0}$ is calculated at five different values of the lattice spacing and then extrapolated to lattice spacing zero. It is well determined just about in the same range as the phenomenological potential. We compare $H(x)$ to the phenomenological potential. the short distance perturbative form and the expected large distance form. The latter is given by an effective bosonic string theory including its universal leading corrections.

Our main motivation, however, was to set the energy scale in the calculation of the running coupling. In ref. [5] a physical coupling was defined for the pure gauge theory in finite volume, that depends only on one scale which is the size of the box $L$. One can calculate the step scaling function which gives the change of the coupling $\bar{g}^2(L)$ when the scale $L$ is changed by a factor $s$. This computation was done in ref. [6] and the lattice results could be extrapolated to the continuum. As a result $\bar{g}^2(L)$ is known for $L/L_8 = 1$, 0.500(23), 0.249(19), 0.124(13) and 0.070(8). Here $L_8$ is defined by $\bar{g}^2(L_8) = 4.765$.

The last step is to calculate the relation of $L_8$ to some physical scale. To this end we previously[6] used the string tension but we could only give a crude estimate of $L_8\sqrt{K}$. In section 4 we will instead calculate $L_8/R_0$ and extrapolate it to the continuum. In the part of the calculation that involves $\bar{g}^2(L)$, we used the Symanzik improved action[5] where the $O(a/L)$ lattice artifacts are removed to one loop order. Comparing this to the results without improvement, we observe a very significant reduction of the lattice artifacts.

## 2.  $R_0$ – a Precise Low Energy Scale

In order to set the scale in a lattice gauge theory calculation, we need to determine a physical dimensionful quantity. As to today's knowledge, the force $F(r)$ between static quarks is the quantity which can be calculated most precisely. In order to avoid an extrapolation in the distance $r$, we define a hadronic length–scale $R(c)$ which depends on a dimensionless parameter $c$ through

$$r^2 \, F(r)|_{r=R(c)} = c \; . \tag{2.1}$$

As discussed in the introduction, choosing e.g. $c = 1.65$ corresponds to

$$R(1.65) \equiv R_0 \simeq 0.5 \text{ fm} \; . \tag{2.2}$$

The advantages of this choice are:

- $R(c)$ is defined precisely, both in the pure gauge theory and in the theory with dynamical fermions. In particular, it does not refer to the quenched approximation.

- In contrast to the string tension it does not contain residual systematic errors that originate from assumptions on subleading terms in the force at intermediate distances.

- It can be calculated with good statistical precision. Therefore its use to set the scale for other physical observables will not introduce significant additional uncertainties into the observables of interest. This is of special importance when one wants to extrapolate lattice results to the continuum. Examples are given in sections 3.2 and 4.

### 2.1  Relationship to other Observables

As in the case of the string tension, one has to use the effective potential in order to relate $R_0$ (or more generally $R(c)$) to experimental observables. Using the Richardson potential, we obtain $R_0 = 0.49$ fm. So e.g. the relationship to the 1P – 1S splitting in the $\bar{c}c$ system $\Delta m$ and to the proton's mass $m_p$ is given by

$$\Delta m \; R_0 = 1.14, \qquad \Delta m \; R(1) = \phantom{0}0.72 \tag{2.3}$$
$$m_p \; R_0 = 2.33, \qquad m_p \; R(1) = \phantom{0}1.47 \tag{2.4}$$

Very similar numbers are obtained when one uses other QCD-inspired phenomenological potentials, whereas the logarithmic potential gives values that are lower by about 10%.

Once one can calculate in full QCD it will be interesting to check eq.'s (2.3,2.4) and one can get a certain measure of the precision of the nonrelativistic description.

## 2.2 Numerical Determination of $R(c)$

It is of vital importance that the quantity that one uses as a reference scale can be computed accurately. We demonstrate that this is the case for $R(c)$ in the following. Three steps have to be discussed in the computation of $R(c)$ in lattice units: 1) the calculation of the potential at a certain distance $\vec{r}/a$, 2) the definition of the force through a finite difference from the potential and 3) the interpolation of the force which is necessary to determine $R(c)/a$ through eq. (2.1).

Step 1) is important but rather technical. It has already been discussed in detail in [7]. Our specific method is described in the appendix. Here, we summarize only the conclusion: using a variational principle and a large enough range of $t$ for the smeared Wilson loops, one obtains an upper bound $V_u(\vec{r})$ and a lower bound $V_l(\vec{r})$ for the potential. We point out that it is advantageous to have larger values of $t$ than the ones used in [7], since then $V_u(\vec{r})$ and $V_l(\vec{r})$ are close to each other and have moderate statistical errors.

Step 2): given the potential, the force at distance $r_I$ and along the orientation $\vec{d}$ is computed from

$$F_{\vec{d}}(r_I) = |\vec{d}|^{-1}[V_l(\vec{r}) - V_l(\vec{r} - \vec{d})], \qquad (2.5)$$

$$r_I = [\, 4\pi \, |\vec{d}|^{-1} \, (G(\vec{r}) - G(\vec{r} - \vec{d})) \,]^{-1/2} \qquad (2.6)$$

$$G(\vec{r}) = a^{-1} \int_{-\pi}^{\pi} \frac{d^3k}{(2\pi)^3} \frac{\prod_{j=1}^{3} \cos(r_j k_j/a)}{4\sum_{j=1}^{3} \sin^2(k_j/2)} \qquad (2.7)$$

Here, the argument $r_I$ is chosen such that $F_{\vec{d}}(r_I)$ is a treelevel improved observable, i.e. to order $g_0^2$ we have $F_{\vec{d}}(r_I) \equiv g_0^2/(4\pi r_I^2)$. This definition removes the $O((a/r)^2)$ lattice artifacts that would be present in the naive choice $r = |\vec{r} - \vec{d}/2|$ as argument of $F$. No fit is necessary to achieve this. Note in addition, that the exact choice of the argument is irrelevant when the force becomes constant. Eq.(2.5) is a convenient definition that greatly reduces the lattice artifacts in the force. We stress, however, that the remaining lattice artifacts have to be controlled by extrapolating simulation results with various (small) $a$ to the continuum. This will be described in the following sections.

We calculated the central values of the force from $V_l$. In order to cover the systematic error of the force that is due to using eq. (A7) at finite values of $t$, we always repeated the calculation of the force with $V_l(\vec{r}) \to V_u(\vec{r})$ and added the resulting difference of the central values to the largest statistical errors of the force. The latter originated from $V_l(\vec{r})$. Although there is no reason that inserting $V_u(\vec{r})$ instead of $V_l(\vec{r})$ into eq. 2.5 will always result in a larger value for the force, this is generally the case in the data. This reflects mainly the fact that the gap eq. (A5) decreases with growing $|\vec{r}|$ and therefore $V_u(\vec{r}) - V_l(\vec{r})$ increases with $|\vec{r}|$.

Step 3) consists of interpolating the force. Here one can simply use the two neighboring points and the interpolation function $F(r) = f_1 + f_2\, r^{-2}$. As this is locally an excellent approximation to the $r$–dependence, the interpolation error is small: it can be estimated from the change that arises from adding a term $f_3\, r^{-4}$ and taking a third point. In the applications discussed below, this change was added as a systematic error. It is well below the statistical uncertainty.

## 3. Reconstruction of the Force in the Continuum

As a first application of setting the scale through $R_0$, we describe here the calculation of the force in the continuum limit.

### 3.1 The Monte Carlo Simulation

In order to be able to extrapolate to the continuum limit, we performed computations at different values of the lattice spacing. We simulated $L^4$ lattices with periodic boundary conditions and the standard Wilson action [8]. According to the state of the art, we used a hybrid algorithm performing $N$ exactly microcanonical overrelaxation sweeps followed by one Creutz heatbath sweep. Table 1 summarizes the parameters used in the simulations. The largest system was simulated on the CERN-IBM, where a speed of about 65 Mflop/s could be achieved.

| $\beta$ | $L/a$ | $S_i$ | $t_{max}/a$ | $M$ | $N$ | $F$ | sweeps |
|---|---|---|---|---|---|---|---|
| 2.50 | 16 | 5, 15, 40 | 8 | 4 | 4 | 1 | 85k |
| 2.55 | 20 | 4, 30 | 9 | 6 | 5 | 2 | 30k |
| 2.60 | 24 | 5, 40 | 10 | 7 | 6 | 3 | 56k |
| 2.70 | 32 | 3, 16, 40 | 8 | 8 | 8 | 2 | 27k |

Table 1: Parameters of our simulations. Measurements were taken every $F \times (N+1)$ sweeps. In order to reduce the auto correlations between measurements, we cycled through the eight sublattices $\Lambda_{\vec{b}}$ (cf. the appendix) from measurement to measurement. $S_i$ gives the number of smearing iterations as discussed in the appendix. $t_{max}$ denotes the maximum time–extent of the Wilson loops and the maximum spatial separation is $r_{max} = M 2\sqrt{3} a$.

Integrated autocorrelation times were estimated for all smeared loops using jacknife binning. They are e.g. roughly 10 sweeps at $\beta = 2.5$ and 14 sweeps at $\beta = 2.6$ for the smallest smeared Wilson loops and *decrease* with increasing loop size.

In table 3, we give the results for the force on all our four lattices. The errors quoted are the jacknife errors[9] on eq. (2.5) plus the change in the central value of the force when we replace $V_l(\vec{r}) \to V_u(\vec{r})$. For $\beta = 2.85$, the corresponding numbers were read off from table 4 of ref. [7]. A number of our $\beta = 2.70$ results can be compared with the ones in [10]. Our central values are all smaller. This difference is somewhat outside of the error bars. Although it is possible that the largest difference is only a typing mistake in [10], it appears that the general difference is due to the smaller range in $t$ used in [10]. The relatively large error bars at $\beta = 2.85$ reflect the fact that the gap in lattice units is small. Hence $V_u(\vec{r}) - V_l(\vec{r})$ becomes significant at such a small lattice spacing and the calculation of the force is difficult at $\beta = 2.85$.

### 3.2 Taking the Continuum Limit

Once we have set the physical scale through $R_0$, the force is entirely described by a dimensionless function:
$$H(x) = F(r)r^2|_{r=xR_0}. \tag{3.1}$$
At different values of the lattice spacing and different orientations $\vec{d_i}$, we obtain the lattice approximations $H_{\vec{d_i}}(x,a)$ to $H(x)$. From Symanzik's discussion of the cutoff dependence of loop integrals[11], one expects that the lattice approximations converge to the continuum with corrections that are roughly proportional to $a^2$:
$$H_{\vec{d_i}}(x, a/R_0) = H(x) + O(a^2/R_0^2) \tag{3.2}$$
The lattice results are plotted in fig. 1 for five sample values of $x$ as a function of the square lattice spacing $[a/R_0]^2$. The extrapolations according to eq. (3.2) are shown as well. Within our precision, no significant $a$–dependence is seen and the extrapolations to the continuum are stable and clearly do not depend on the functional form of the assumed $a$–dependence. To a large extent this is due to the tree level improved definition of the force (cf. section 2.2).

Extrapolations were done for a range of $0.35 \leq x \leq 2.1$. At smaller values of $x$, we have not performed the extrapolation since there we have only two points in $a$. For $x > 1.1$ there is data with $\beta \leq 2.7$ only.

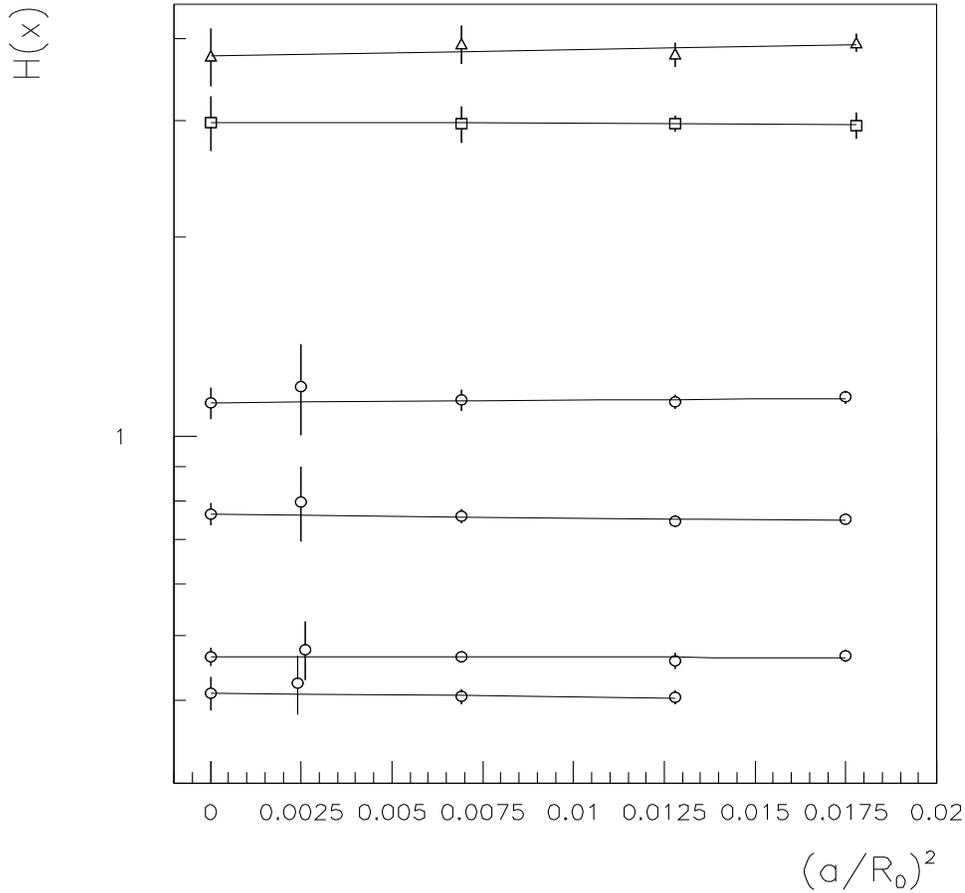

Figure 1: Continuum extrapolation for $H(x)$ with $x = 1.6$ ($\triangle$) $x = 1.4$ ($\square$), and $x = 0.8, 0.6, ,0.4, 0.35$ ($\circ$) from top to bottom. The circles are for the force along orientation $\vec{d}_1$, the squares correspond to orientation $\vec{d}_2$ and the triangles to orientation $\vec{d}_3$.

### 3.3  Discussion of the Force

The continuum results for $H(x)$ are plotted in fig. 2. It is determined accurately in about the same distance range, where the universality of the phenomenological force holds[4].

We compare this SU(2) Yang–Mills force with the phenomenological model by Richardson[3]. As can be seen e.g. in fig. 1 ref. [4], other phenomenological models do not differ much from the Richardson model in the range of $x$ discussed here. We note, however, that the force of the Martin[12] and the logarithmic potential [13] differs from the Richardson force as much as the SU(2) Yang–Mills force. In this sense, even without fermions and for gauge group SU(2), the static force compares quite well with the phenomenological forces.

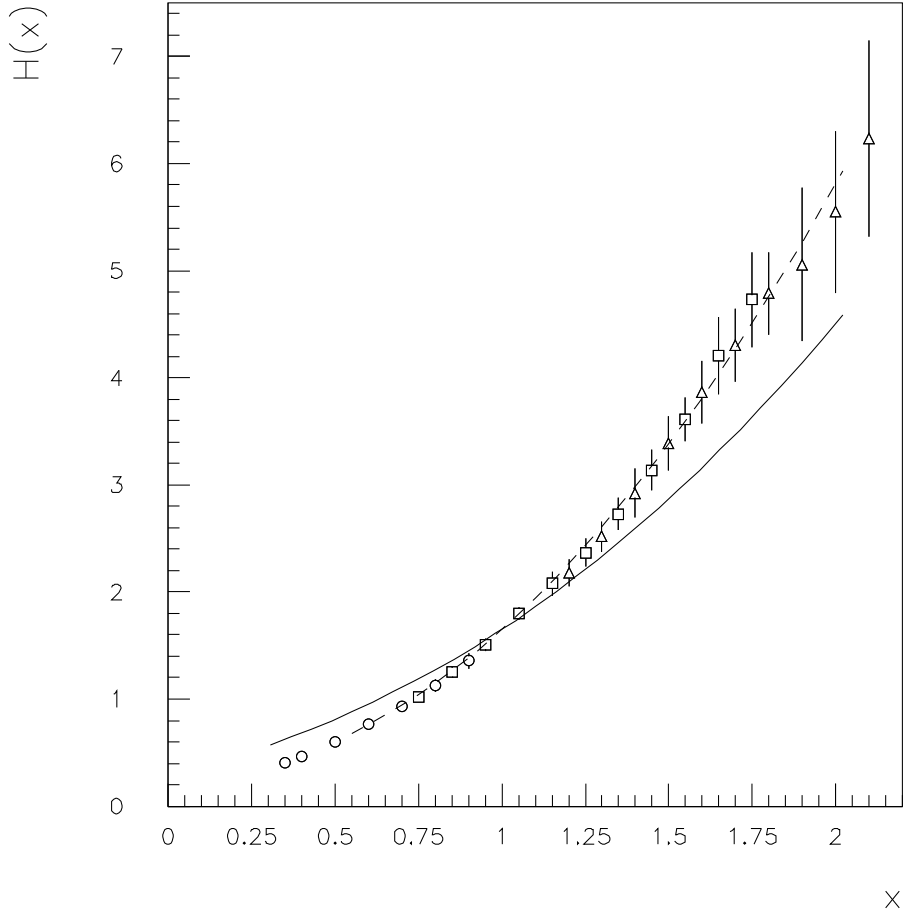

Figure 2: $H(x) = F(r)r^2|_{r=xR_0}$ as a function of $x$. Symbols as in fig. 1. The Richardson model is shown as full line and the bosonic string model (normalized by $H(1) \equiv 1.65$) as dashed curve.

In the figure, we also show the prediction of the bosonic string model normalized at $x = 1$. It includes besides the string tension the universal $\pi/(12r^2)$ correction to the force. This correction should be there at large $r$ [14]. As already noted earlier[15], this form gives an excellent *effective* representation of the force also for rather small values of $x$. Note that previous claims in the literature[16], that the Yang-Mills Wilson loops are effectively described by a *fermionic* string model are based on the subtraction of perturbative contributions.

### 3.4 The Running Coupling $\alpha_{q\bar{q}}(r)$

The quantity $H(x)$ may be used to define a physical running coupling

$$\alpha_{q\bar{q}}(r) = \frac{4}{3}H(x), \quad r \equiv xR_0. \tag{3.3}$$

It is of interest to see, in how far we have reached the perturbative region in $x$. To check this, we start with our value of $\alpha_{q\bar{q}}(r)$ at the smallest value of $r$, and integrate the 1-loop and the 2-loop perturbative renormalisation group equations towards larger $r$. This is shown in fig. 3. We see, that the 2-loop $\beta$-function describes the evolution of the coupling for only a change in scale up to a factor two. In addition, the 1-loop running is very different, since the 2-loop term in the $\beta$-function contributes 25% and more in this range of $r$. We conclude that $r \simeq 0.17$ fm is outside of the universal perturbative domain.

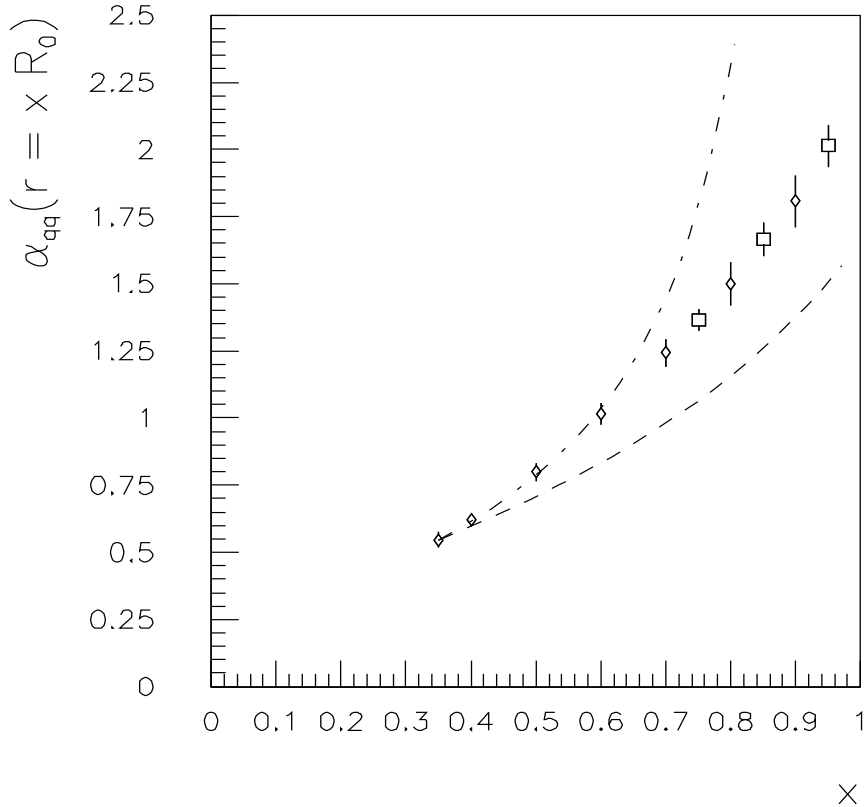

Figure 3: $\alpha_{q\bar{q}}(r)$ together with 1-loop (dashed) and 2-loop (dotted-dashed) renormalisation group evolution starting at the smallest value of the coupling.

We may also consider the perturbative relation between $\alpha_{q\bar{q}}(r)$ and the finite volume coupling $\bar{g}^2(L)$ that will be discussed further in the following section. This reads $\alpha_{q\bar{q}}(r) = \alpha(L) + 0.9980\, \alpha^2(L) + ...$, $\alpha(L) \equiv \bar{g}^2(L=r)/(4\pi)$. Using the result of the following section, we get $\alpha_{q\bar{q}}(r = 0.35 R_0) = 0.39 + O(\alpha^3)$, which is significantly below the nonperturbative result $\alpha_{q\bar{q}}(r = 0.35R_0) = 0.54(2)$. However, at this value of $r$, the $O(\alpha^2)$ term is 30% of the $O(\alpha)$ term and we arrive again at the conclusion that $r$ is too large to use perturbation theory. Within our approach of extrapolating the lattice numbers to the continuum, we could not reach significantly smaller values of $r$.

## 4. Setting the Physical Scale in the Computation of the Running Coupling $\alpha(L)$

In ref. [6], a renormalised coupling $\bar{g}^2(L)$, that runs with the box–size $L$ was calculated for different length scales $L$.[1] At the largest scale investigated in ref. [6], the coupling has the value $\bar{g}^2(L) = 4.765$. At this value one can make contact with the low energy scales of the theory in infinite volume: one calculates the product $LE$, with $E$ some energy scale of the theory in infinite volume. The choice of ref. [6] was $E = \sqrt{K}$, with $K$ the string tension, since data for two small values of the lattice spacing existed for this quantity. This can be improved by using $E = 1/R(c)$ instead.

We describe now the computation of $L/R(c)$, extrapolated to the continuum limit. In a first step, we determine the bare coupling $\beta$ as a function of the lattice size $L/a$ for fixed $\bar{g}^2(L) = 4.765$. Then, we use $R(c)/a$ at different $\beta$–values in the same range, form the ratio $L/R(c)$ and extrapolate it to the continuum $a/L = 0$.

| $L/a$ | $\beta$ | $\beta_{impr}$ |
|---|---|---|
| 5 | 2.4971(23) | 2.5500(23) |
| 6 | 2.5752(28) | 2.6082(30) |
| 7 | 2.6376(20) | 2.6674(24) |
| 8 | 2.6957(21) | 2.7155(21) |
| 10 | 2.7824(22) | 2.7989(23) |
| 12 | 2.8485(32) | 2.8697(39) |
| 14 | 2.9102(62) | |

Table 2: Bare coupling $\beta$ from ref. [6] and bare coupling $\beta_{impr}$ for the 1–loop improved action vs. the lattice size at fixed $\bar{g}^2(L) = 4.765$.

At the largest values of the coupling, a significant lattice spacing dependence of the step scaling function was found in ref. [6] (cf. fig. 2 in that reference). It was therefore to be expected, that $L/R(c)$ shows lattice artifacts. We attempted to reduce these artifacts by using the 1–loop Symanzik improved action given in ref. [5]. We denote the results that are obtained using this action with a subscript "$impr$". The $\beta$–values are listed in table 2 together with the ones without improvement that were already obtained in ref. [6].

---
[1]For details on the definition and the computation of $\bar{g}^2(L)$, we refer the reader to [5, 6].

Both $\beta(L/a)|_{\bar{g}^2(L)=4.765}$ and $\beta_{impr}(L/a)|_{\bar{g}^2(L)=4.765}$ are almost linear functions of $\log(L/a)$. Therefore, the (noninteger) values $L/a$ and $L_{impr}/a$ for the values of $\beta$ where we know the force (table 3) are easily determined by interpolation. At these $\beta$–values values, we calculated $R_{\vec{d}_i}(c)/a$ as discussed in section 2.2.

Examples of the data for $L/R_{\vec{d}}(c)$ are plotted in fig. 4. It is well worth noting that fig. 4 is the first time that the success of the Symanzik improvement program can be demonstrated in 4–dimensional pure gauge theories: the dependence of the ratios on the lattice spacing is much weaker when the 1–loop improved action is used. In fact, due to the weak cutoff dependence, the data with improvement can be extrapolated to the continuum with confidence.

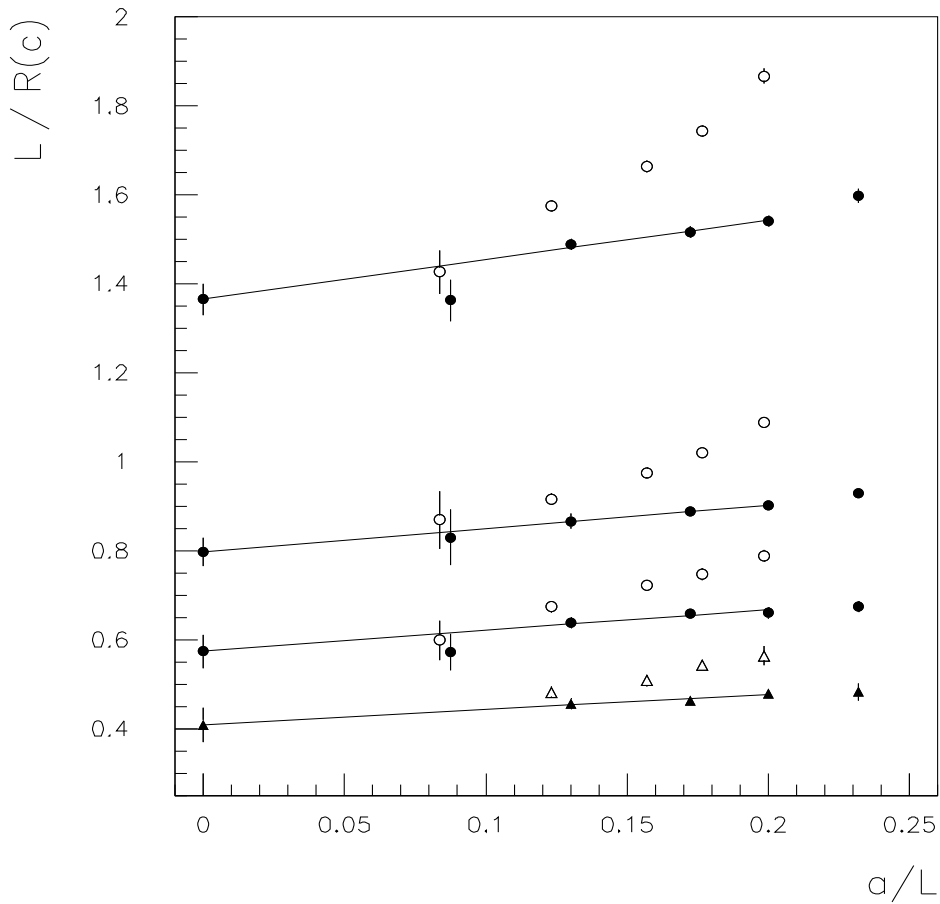

Figure 4: $L/R_{\vec{d}_1}(c)$ (○), $L_{impr}/R_{\vec{d}_1}(c)$ (●), $L/R_{\vec{d}_3}(c)$ (△) and $L_{impr}/R_{\vec{d}_3}(c)$ (filled △) as function of the lattice spacing $a/L$ and $c = 0.5$, 1.0, 1.65 and 3.0 from top to bottom. The lines are the linear extrapolation to the continuum using $a/L_{impr} \leq 1/5$. At $a/L_{impr} = 0$ the extrapolated values are plotted together with their error bars.

The physically most interesting case is around $c = 1.65$ and we use $R(1.65) = R_0 \simeq 0.5$ fm in the following (at finite values of the lattice spacing, $R_{\vec{d}}(c)$ depends on the orientation $\vec{d}$. In the continuum limit, this dependence disappears and we drop the corresponding index.) Our extrapolation to the continuum yields

$$L/R_0 = 0.573(37) \qquad (4.1)$$

or — using $R_0 = 0.49$ fm as an illustration also for SU(2) Yang–Mills theory — $L = 0.281(18)$fm. For comparison, we also give the numbers that were obtained from the force along the other two orientations: $L/R_0 = 0.587(39)$ for $\vec{d}_2$ and $L/R_0 = 0.595(31)$ for $\vec{d}_3$. The latter extrapolations had to be performed without the $\beta = 2.85$ points.

These extrapolations are stable: Within the error bars, we obtain the same results if we include one more point at larger $a$ or if we remove the last point from the extrapolation. Furthermore, the difference of the extrapolated value from the point at largest $a$ that was included in the extrapolation is only about 15%. Therefore, the above calculation gives the momentum–scale for the running coupling $\bar{g}^2(L)$ to a precision of about 6% in the continuum limit.

It is more easy to control systematic errors in the calculation of $R_0$ than in the determination of the string tension. Nevertheless, it is interesting to check, in how far one obtains a different result if the scale is set by the string tension. We have therefore determined the string tension from $F_{\vec{d}_i}(r) = K + \pi/(12r^2)$ for $r^2 K \geq 1.8$. This range in $r$ was chosen such that the $\beta = 2.85$ data could be included. The $\pi/(12r^2)$ correction[14] is less than 14% and the results of section 3.3 provide some evidence that the uncertainty on the correction may be neglected. Extrapolating $L_{impr}\sqrt{K}$ to the continuum as shown in fig. 5, gives

$$L\sqrt{K} = 0.675(60) \qquad (4.2)$$

or (with $\sqrt{K} = 425$MeV) $L = 0.313(28)$fm. The small change compared to the determination that uses $L/R_0$ is due to the fact that the force in SU(2) Yang–Mills theory deviates somewhat from the Richardson model in the relevant $r$–range.

The statistical uncertainty in both determinations of $L$ in physical units is quite similar, because they originate from the force in about the same range of $r$. However, the string tension determination uses the assumption on the subleading correction to the force. Although fig. 2 provides reasonable evidence that we may use the correction derived from bosonic string vibrations, an uncertainty due to this assumption is not easy to quantify.

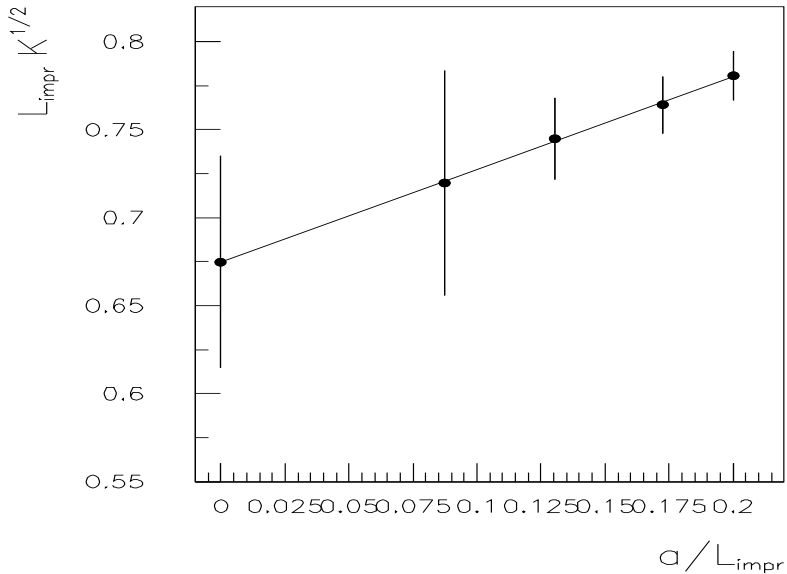

Figure 5: The extrapolation of $L_{impr}\sqrt{K}$ to the continuum limit $a/L_{impr} = 0$.

## 5. Conclusions

It has already been pointed out in [6], that a good way to set the scale in the pure gauge theory is through the force at some finite distance well inside the nonperturbative region. We have shown here that this proposal can be carried out in practice and that the quantity $R_0$ provides a low energy scale that can be calculated precisely. Compared to the string tension this avoids the extrapolation in $r$ and is therefore independent of parametrisations of the force. The relation to experiment is based on the assumption of identifying the static QCD potential and the phenomenological potential.

Using $R_0$ and the 1–loop improved action for the simulation determining $\bar{g}^2(L)$, we were able to compute the scale in the running coupling of [6, 5] to within 6% in the continuum limit.

The force itself could be extrapolated to the continuum limit with all systematic errors taken into account. In order to do this, it was most important to have precise data at *several* rather small values of the lattice spacing. In addition, lattice artifacts could be suppressed by choosing a tree–level improved finite difference for the lattice force in terms of the lattice potential.

$R_0$ calculated in the way described here will clearly be a useful reference scale also in the SU(3) pure gauge theory. It should be mentioned, however, that once one approaches full QCD, there are other quantities that are *directly* measured in experiments that can play this role. Nevertheless it will be interesting to reconstruct the force along the lines of section 3.2 and compare to phenomenological models. In addition, a check of the relations eq.'s (2.3,2.4) will be of interest.

**Acknowledgement:** I thank U. Wolff for the collaboration calculating the data in Table 2. I have profited from a number of discussions with M. Lüscher and I thank him in particular for critically reading the manuscript.
The most important numerical computations of the force were performed on the IBM at CERN. I would like to thank the CERN-CN staff as well as Pierre Aubry from IBM for their excellent support.

## A  The Calculation of the Potential

The force between static quarks is most efficiently calculated from "smeared" Wilson loops. We have applied a variation of smearing based on the original ideas of ref. [18, 19]. The smearing and variational technique employed her deviates only slightly from the one used by Michael et al. [10, 7]. A detailed discussion of the merits of the method can be found in [7]. Here, we describe the exact implementation that was used and discuss the results.

We start from lattice gauge fields $U_\mu(x) \in \mathrm{SU}(2)$, that are the gauge connections between points $x$ and $x + \hat{\mu}a$, $\mu = 0, 1, 2, 3$ of a hypercubic lattice $\Lambda$ with lattice spacing $a$. In a first step, the space–components of the gauge field are "smeared" by one iteration

$$U_k(x) \to U'_k(x) = \mathcal{P} \left\{ U_k(x) + \omega \sum_{j \neq k=1}^{3} [\, U_j(x)U_k(x+\hat{j}a)U_j^\dagger(x+\hat{k}a) \right. \tag{A1}$$
$$\left. + \; U_j^\dagger(x-\hat{j}a)U_k(x-\hat{j}a)U_j(x-\hat{j}a+\hat{k}a) \,] \right\}, \quad k = 1, 2, 3,$$

Here $\mathcal{P}$ denotes the projection into SU(2) and we have used $\omega = 1/4$. Next we block to a sub–lattice $\Lambda_{\vec{b}}$ characterized by a vector $\vec{b}$ with components $b_i = 0, 1$. The sublattice has $1/2$ the number points per space–dimension by restricting the components of $\vec{x}/a$ to be even ($b_i = 0$) or odd: $x_i/a = b_i$ mod 2, $i = 1, 2, 3$ for $x \in \Lambda_{\vec{b}}$. The gauge fields on the sublattice are simply

$$\tilde{U}_k(x) = U_k(x)U_k(x+\hat{k}a), \quad x \in \Lambda_{\vec{b}} \,. \tag{A2}$$

On this blocked lattice we apply eq. (A1) $S_i$ times (with $a \to 2a$ and $U \to \tilde{U}$). In order to be able to calculate the potential along the three orientations $\vec{d}_1 = 2a\,(1,0,0)$, $\vec{d}_2 = 2a\,(1,1,0)$, $\vec{d}_3 = 2a\,(1,1,1)$, we then construct generalized gauge fields $V^{(i)}$ that connect $x = (x_0, \vec{x})$ with $(x_0, \vec{x} + \vec{d}_i)$, $i = 1, 2, 3$. $V^{(i)}$ are obtained by averaging the parallel transporters over the different shortest lattice paths on $\Lambda_{\vec{b}}$.

We computed smeared Wilson loops with space extent $\vec{r} = n_i \vec{d_i}$, $n_i = 1, 2, ..., M$ using the appropriate product of $V^{(i)}$ for the space–like parts and the 1-link integral[20] for the time–like parts of the loops. Inserting $V$ after smearing level $S_i$ at time 0 and $S_j$ at time $t$, one obtains a matrix correlation function. The latter corresponds to matrix elements $(\hat{T}^n(\vec{r}))_{ij}$, $n \equiv t/a$ of powers of the transfer matrix $\hat{T}$ in the corresponding charged sector of the Hilbert space.[2]

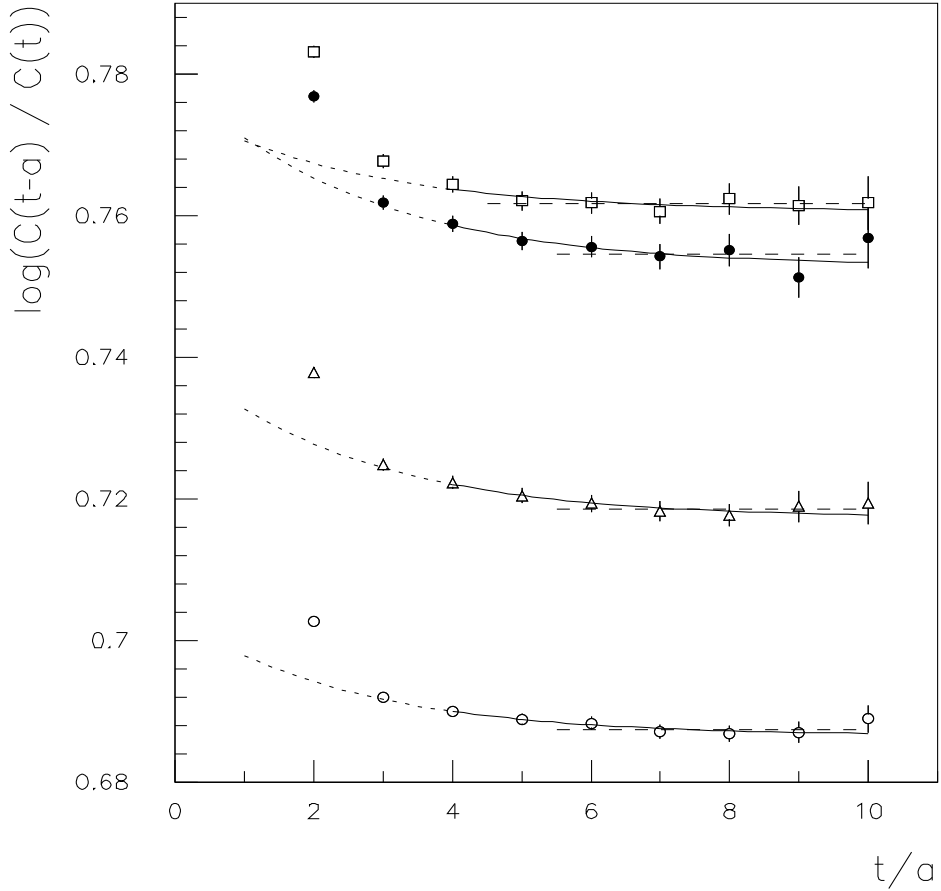

Figure 6: $\log[C(t,\vec{r})/C(t-a,\vec{r})]$ at $\beta = 2.6$. The data points are from top to bottom for $\vec{r} = 5\vec{d_2}$, $4\vec{d_3}$, $6\vec{d_1}$ and $3\vec{d_3}$. The long dashes represents the fit to eq. (A6) and the full line is the fit to eq. (A7), plotted in their respective fit ranges. As an illustration, the fits to eq. (A7) are continued to $t-$ values outside of the fit ranges as short dashed curve.

The generalized eigenvalue equation

$$\sum_j [\hat{T}^{n+1}(\vec{r})]_{ij} v(\vec{r})_j^{(k)} = \lambda_k(\vec{r}) \sum_j [\hat{T}^n(\vec{r})]_{ij} v(\vec{r})_j^{(k)}, \quad \lambda_1(\vec{r}) \leq \lambda_2(\vec{r}), ... \quad \text{(A3)}$$

---

[2]In this discussion we neglect corrections due to a finite time extent of the lattice and use the convention that the energy of the vacuum is zero.

gives estimates $\lambda_k(\vec{r})$ that converge, for $n \to \infty$, (exponentially) to the lowest eigenvalues of the transfer matrix[21].

In our simulations (cf. section (3.1)), we found that the first two eigenvalues are fairly stable with respect to $n$, once $n \geq 2$. Higher eigenvalues are difficult to determine since – apparently – the different states generated by the above smearing procedure are not linearly independent to a sufficient degree. We used this variational technique to construct a correlation function

$$C(t,\vec{r}) = \sum_{i,j} v(\vec{r})_i^{(1)} [\hat{T}^n(\vec{r})]_{ij} v(\vec{r})_j^{(1)} \tag{A4}$$

that obtains only small contributions from excited states because it is projected onto the approximate ground state. Furthermore, we estimated the gap in lattice units by

$$a\,\Delta(\vec{r}) = \log(\lambda_1(\vec{r})/\lambda_2(\vec{r})). \tag{A5}$$

We then determined the potential from fits

$$C(t,\vec{r}) = C_1^2(\vec{r}) \exp(-V_u(\vec{r})t) \text{ and} \tag{A6}$$

$$C(t,\vec{r}) = C_1^2(\vec{r}) \exp(-V_l(\vec{r})t)[1 + \frac{C_2^2(\vec{r})}{C_1^2(\vec{r})} \exp(-\Delta(\vec{r})t)]. \tag{A7}$$

Here $\Delta(\vec{r})$ was fixed to the value determined from the variational method eq. (A5). $V_u(\vec{r})$ is an upper bound for the potential since the excited state contributions are neglected in eq. (A6). On the other hand, $V_l(\vec{r})$ gives a lower bound for the potential provided $\Delta(\vec{r})$ as determined from eq. (A3) is a reasonable estimate for the gap. This is true, because eq. (A7) parametrizes all corrections to the lowest state by an effective contribution from the first excited state. It thus overestimates the corrections at large $t$ (i.e. $t$–values outside of the fit range). As they are lower and upper bounds, one can – in principle – use the combination of eq. (A6,A7) for any fit ranges (in $t$), irrespective of whether the fits are statistically satisfactory. If this is done allowing too small values of $t$, however, $V_u(\vec{r})$ and $V_l(\vec{r})$ become significantly different and one has a dominant systematic error $V_u(\vec{r}) - V_l(\vec{r})$. In the present calculation, we have a relatively large range in $t$. We chose the fit ranges such that they yield a satisfactory $\chi^2$ (including the correlations of the data). This resulted in $V_u(\vec{r}) = V_l(\vec{r})$ within one standard deviation.

We give examples of the data together with fits in the form of the time derivative of $C(t,\vec{r})$ in fig. 6. The values of $\vec{r}$ chosen for the figure are in the range that is most relevant for the analysis of the force discussed in this work. It is evident from the figure that $C(t,\vec{r})$ provides a correlation function which receives only small contributions from excited states in the accessible range of $t$. Therefore we can extract the ground state potential with moderate error bars.

| $\beta$ | $r_I/a$ | $F_{\vec{d}_1}(r_I)a^2$ | $r_I/a$ | $F_{\vec{d}_2}(r_I)a^2$ | $r_I/a$ | $F_{\vec{d}_3}(r_I)a^2$ |
|---|---|---|---|---|---|---|
| 2.50 | 2.654 | 0.0698( 3) | 4.008 | 0.0506( 2) | 4.960 | 0.0451( 3) |
| 2.50 | 4.794 | 0.0458( 3) | 6.939 | 0.0395( 7) | 8.517 | 0.0380( 9) |
| 2.50 | 6.865 | 0.0395(11) | 9.807 | 0.0374(10) | 12.02 | 0.0357(26) |
| 2.55 | 2.654 | 0.0588( 2) | 4.008 | 0.0401( 2) | 4.960 | 0.0351( 3) |
| 2.55 | 4.794 | 0.0351( 4) | 6.939 | 0.0303( 4) | 8.517 | 0.0280( 6) |
| 2.55 | 6.865 | 0.0299( 5) | 9.807 | 0.0271( 8) | 12.02 | 0.0274(10) |
| 2.55 | 8.899 | 0.0275(10) | 12.65 | 0.0263(13) | 15.50 | 0.0277(23) |
| 2.55 | 10.91 | 0.0266( 9) | 15.49 | 0.0255(20) | 18.98 | 0.0247(38) |
| 2.60 | 2.654 | 0.0507( 1) | 4.008 | 0.0333( 1) | 4.960 | 0.0280( 2) |
| 2.60 | 4.794 | 0.0286( 2) | 6.939 | 0.0230( 3) | 8.517 | 0.0209( 5) |
| 2.60 | 6.865 | 0.0229( 4) | 9.807 | 0.0205( 3) | 12.02 | 0.0192( 6) |
| 2.60 | 8.899 | 0.0212( 3) | 12.65 | 0.0193( 5) | 15.50 | 0.0187( 8) |
| 2.60 | 10.91 | 0.0210(17) | 15.49 | 0.0192( 8) | 18.98 | 0.0187( 6) |
| 2.60 | 12.93 | 0.0196( 6) | 18.33 | 0.0193( 8) | 22.46 | 0.0196(14) |
| 2.70 | 2.654 | 0.0406( 1) | 4.008 | 0.0245( 1) | 4.960 | 0.0197( 1) |
| 2.70 | 4.794 | 0.0201( 1) | 6.939 | 0.0149( 2) | 8.517 | 0.0132( 2) |
| 2.70 | 6.865 | 0.0151( 1) | 9.807 | 0.0122( 3) | 12.02 | 0.0114( 3) |
| 2.70 | 8.899 | 0.0127( 3) | 12.65 | 0.0113( 3) | 15.50 | 0.0105( 5) |
| 2.70 | 10.91 | 0.0117( 4) | 15.49 | 0.0104( 6) | 18.98 | 0.0107( 5) |
| 2.70 | 12.93 | 0.0112( 4) | 18.33 | 0.0106( 5) | 22.46 | 0.0101( 7) |
| 2.70 | 14.94 | 0.0106( 5) | 21.17 | 0.0105( 5) | 25.93 | 0.0101( 6) |
| 2.85 | 2.654 | 0.0318( 2) | | | | |
| 2.85 | 4.794 | 0.0137( 3) | | | | |
| 2.85 | 6.865 | 0.0089( 5) | | | | |
| 2.85 | 8.899 | 0.0067( 4) | | | | |
| 2.85 | 10.91 | 0.0056( 7) | | | | |
| 2.85 | 12.93 | 0.0055( 4) | | | | |
| 2.85 | 14.94 | 0.0050( 7) | | | | |
| 2.85 | 16.94 | 0.0044( 3) | | | | |
| 2.85 | 18.95 | 0.0037(10) | | | | |
| 2.85 | 20.95 | 0.0045( 6) | | | | |
| 2.85 | 22.96 | 0.0045( 7) | | | | |

Table 3: Force and improvement radius in lattice units. For convenience we have also included the results of [7]. In physical units $r_I$ covers roughly the same range for all five $\beta$.